\documentclass[12pt]{JHEP3}
\usepackage{epsfig}
\usepackage{amsmath}
\newcommand{\unit}{\hbox to 3.8pt{\hskip1.3pt \vrule height 7.4pt
    width .4pt \hskip.7pt \vrule height 7.85pt width .4pt \kern-2.4pt 
    \hrulefill \kern-3pt \raise 3.7pt\hbox{\char'40}}}
\newcommand{\p}{\partial}

\newcommand{\nn}{\nonumber}

\title{
Reconnection of Colliding Cosmic Strings
}
\author{
Amihay Hanany$^*$ and Koji Hashimoto$^\dagger$\\ 
$^*$Center for Theoretical Physics, 
Massachusetts Institute of Technology,\\
Cambridge, MA 02139, U.S.A.\\
E-mail: \email{hanany@mit.edu}\\
$^\dagger$Institute of Physics, University of Tokyo\\
Komaba, Tokyo 153-8902, Japan\\
E-mail: \email{koji@hep1.c.u-tokyo.ac.jp}\\
}

\abstract{
For vortex strings in the Abelian Higgs model and D-strings in
superstring theory, both of which can be regarded as cosmic strings, 
we give analytical study of reconnection (recombination,
inter-commutation) when they collide, by using effective field theories 
on the strings. First, for the vortex strings, via a string sigma model,
we verify analytically that the reconnection is classically inevitable
for small collision velocity and small relative angle. Evolution of the
shape of the reconnected strings provides an upper bound on the
collision velocity in order for the reconnection to occur. These
analytical results are in agreement with previous numerical results.
On the other hand, reconnection of the D-strings is not classical but
probabilistic. We show that a quantum calculation of the reconnection
probability using a D-string action reproduces the nonperturbative
nature of the worldsheet results by Jackson, Jones and Polchinski. 
The difference on the reconnection --- classically inevitable for the
vortex strings while quantum mechanical for the D-strings --- is
suggested to originate from the difference between the effective field 
theories on the strings. 
}

\preprint{
{\normalsize{\tt hep-th/0501031}}\\ 
{\normalsize MIT-CTP-3584}\\
{\normalsize UT-Komaba/05-1}
}

\begin{document}

\section{Introduction}
\label{sec1}

Recent revival of the study of cosmic strings \cite{Kibble} originates
partially in the proposal that cosmic strings can be fundamental
superstrings, D-strings, or their bound states called $(p,q)$-strings 
in various compactification scenarios \cite{Tye,CMP}.\footnote{See also
reviews \cite{reviews}.} Many other options like wrapped branes on
supersymmetric cycles or tensionless strings can be good candidates as
well. Though first mostly denied by E.~Witten \cite{Witten} 20 years
ago, this possibility was re-born thanks to the decade's developments in
string/M-theory which have provided fertile bases for constructing
semi-realistic universes in terms of D-branes. Pursuing the possibility
is quite important by the obvious reason that we may observe some
signals of those very stringy objects directly in the sky.  

\begin{figure}[tp]
\begin{center}
\begin{minipage}{14cm}
\begin{center}
\includegraphics[width=12cm]{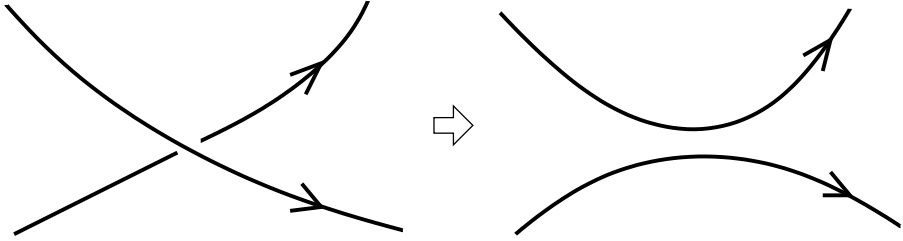}
\caption{Reconnection of strings.}
\label{fig1}
\end{center}
\end{minipage}
\end{center}
\end{figure}

For cosmic strings, 
various classical solutions such as vortex strings in Abelian Higgs
models and other scalar/gauge field theories allowing
topological vortices have been adopted to model them and study 
their properties.\footnote{Note that Type I/II cosmic strings are
not related to Type I/II superstrings. The former types are
distinguished by a relation between the two coupling constants in the
Abelian Higgs model. In addition, ``gauge strings'' refer to vortex
strings in the Abelian Higgs model, while ``global strings'' arise
in scalar field theories with topologically nontrivial vacua.} 
One of the points which distinguishes the fundamental strings or the 
D-strings from the field theory vortex strings is in their {\it
reconnection probability} \cite{CMP}. When strings collide with a
relative angle, they may be cut once at the collision and connected at
the different ends. This is called reconnection (or recombination,
inter-commutation), see Fig.~\ref{fig1}. This reconnection probability
is one of the indispensable ingredients for simulating galaxy formation
in the early universe, and is important also for the direct detection of 
gravitational waves arising from cusps created when the cosmic strings
are reconnected. For the vortex strings, numerical simulations
\cite{Shellard, Matzner, Moriarty} have been extensively performed and
exhibit the universal feature: {\it these vortex strings always
reconnect classically for small collision velocities, while above a
velocity upper bound they do not reconnect.} On the other hand, {\it for
fundamental strings and D-strings, the reconnection is probabilistic for
any collision velocity} \cite{JJP}. Based on the difference between the
two, some numerical simulations depending on this probability were
reported recently \cite{report}. 

In this paper, we clarify the origin of this difference theoretically
and analytically. In a word, it comes from the difference of the 
effective field theories on the strings. For the effective action of
the D-strings (which is called a D-string (or D-brane) action)
there are classical solutions representing D-strings passing through
each other without reconnection. Since such classical solutions exist,
the reconnection occurs only in a probabilistic manner.\footnote{
For fundamental strings, the reconnection is probabilistic from the
first place as known in an interaction in a string field theory 
\cite{HIKKO}. So we don't deal with the fundamental strings in this
paper.} 
On the other hand, for vortex strings the effective field
theory on those does not allow such classical solutions and
thus the reconnection takes place classically, and inevitably. 

The effective field theory on the vortex is a sigma model whose target
space has been derived recently by using brane realization techniques
\cite{HT}. Thus it comes from a D-brane action, but the presence of
additional matter fields due to the specific brane configuration and
also the field theory limit make the resulting effective field theory of
the vortex strings look very different from the usual D-brane
action. More precisely, the theory on the vortex string contains
fundamental matter fields and the relevant dynamics is in the Higgs
phase of the effective theory. We give an analytical proof that the
vortex strings always reconnect for small collision velocity and
collision angle, by using the effective sigma model. This technique also
allows to explain the existence of the velocity upper bound, and we
derive it analytically by considering a geometrical constraint on the
shape of the reconnected vortex strings. It turns out that the upper
bound coincides with the result of \cite{CT} which was obtained by
looking at deformations of classical solutions of the Abelian Higgs
model. 

For colliding D-strings, classically nothing occurs because the usual 
D-string action allows a simple solution describing the D-strings
passing through each other without reconnection. But quantum
mechanically there appears a tachyonic instability intrinsic to
intersecting D-branes \cite{BDL, Aki-Wati}. The tachyon condensation
leads to the reconnection \cite{HN}, and we study time evolution of
a tachyon wave function to evaluate the reconnection probability.
The result shares the same non-perturbative property as found in the
string worldsheet calculation in \cite{JJP}.

The organization of this paper is as follows. In Section \ref{sec2}, we 
show two effective actions and give the classical difference between the 
colliding solutions, to explain the difference between the reconnection
property of the D-strings and the reconnection property of the vortex 
strings. In Section \ref{sec3},
we give the analytical proof of the reconnection of the vortex strings
and the velocity upper bound. In Section \ref{sec4}, the D-string
reconnection probability is calculated. Section \ref{sec5} is devoted to
a summary and discussions.


\section{Vortex strings and D-strings}
\label{sec2}

The Abelian Higgs model of our interest can be realized as a 
field theory limit of a theory on a D4-brane embedded in a
certain brane configuration \cite{Hanany:1996ie}, 
and therefore the vortex strings can also be described as a D-brane
in that brane configuration \cite{HT}. We shall review the
D-brane action and the brane configuration, in order to clarify the 
difference between the vortex strings and the D-strings themselves.

First, we consider D-strings in the absence of other kinds of branes. 
The D-strings are moving in 10 dimensional spacetime, and in some
compactification scenario these D-strings can be thought of as cosmic
strings. The bosonic part of the D-string low energy 
action in flat target spacetime is
\begin{eqnarray}
 S =  \frac{2\pi l_{\rm s}^2}{g_{\rm s}}
\int\! dt dx \;{\rm Tr}\left[-\frac{1}{4}
F_{\mu\nu} F^{\mu\nu} - \frac{1}{2}D_\mu \Phi_i D^\mu \Phi^i
+ \frac{1}{4}[\Phi_i,\Phi_j]^2
\right]
\label{effe}
\end{eqnarray}
where the eigenvalues of $2\pi l_{\rm s}^2\Phi$ ($l_{\rm s}$ is the
string length) measures the target space location transverse to the
worldsheet of the D-strings, and thus $i$ runs from 2 to 9, the
transverse dimensions. We consider a pair of D-strings tilted and
colliding with each other, and in fact there is a classical solution
representing them passing through each other without reconnection:
\begin{eqnarray}
 2\pi l_{\rm s}^2\Phi_2 = \left(
\begin{array}{cc}
\tan (\theta/2)x& 0  \\ 0 & -\tan(\theta/2) x
\end{array}
\right) \ , \quad 
 2\pi l_{\rm s}^2 \Phi_3 = \left(
\begin{array}{cc}
\overline{v}t& 0  \\ 0 & -\overline{v}t 
\end{array}
\right) \ .
\label{passing}
\end{eqnarray}
Here $\theta$ is the relative angle between the D-strings, and 
$2\overline{v}$ is the relative velocity.\footnote{
The action (\ref{effe}) is valid for a slow motion of the D-strings.
To treat fast (relativistic) collisions, one has to use the
non-Abelian Born-Infeld action.  } At the collision incidence, nothing
like the reconnection occurs. 
The solution represents two straight D-strings, but any collision of
D-strings can be described by this solution at least locally around the
collision point. This immediately tells us that colliding D-strings do
not cause the reconnection classically.\footnote{However, a quantum
treatment shows probabilistic reconnection via condensation of tachyonic
fundamental strings connecting the two D-strings, as we will see in
Section \ref{sec4}. } This argument is valid as long as the D-string
action is effective: for decoupling of closed strings 
($g_{\rm s}\rightarrow 0$) and at the low energy ($\theta \ll 1$ and
$\bar{v} \ll 1$ for the configuration of (\ref{passing})).  

On the other hand, we claim that any collision of vortex strings in the
Abelian Higgs model causes the reconnection if the collision velocity
and the collision angle are small enough, that is, for the same
parameter region ($\theta \ll 1, \bar{v}\ll 1$). We first derive the
effective theory on the pair of the vortex strings by following
\cite{HT}, then show that any classical solution of the form 
(\ref{passing}) does not exist. The derivation in \cite{HT} is through
brane configurations in Type IIA string theory, and resultantly the
effective action comes from a D-brane action and thus resembles
(\ref{effe}). However, as we will see, there appears a crucial 
difference.  

\begin{figure}[tp]
\begin{center}
\begin{minipage}{14cm}
\begin{center}
\includegraphics[width=12cm]{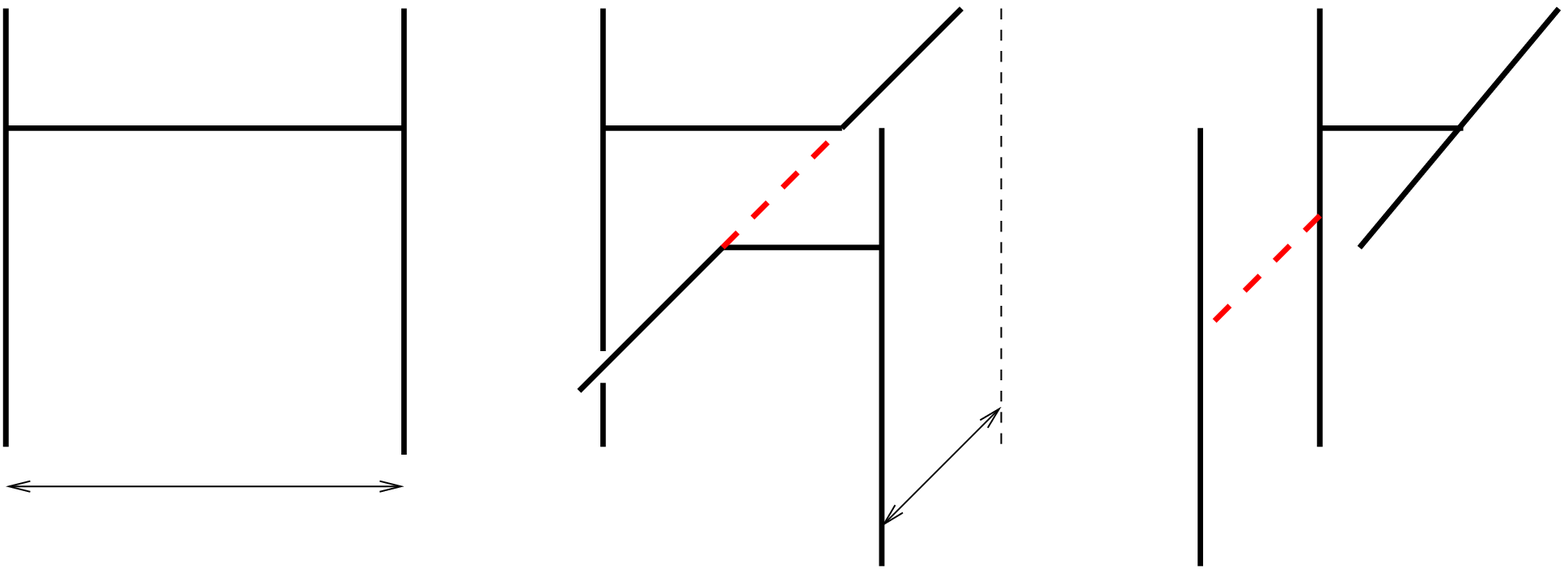}
\put(-350,125){NS5}
\put(-265,125){NS5}
\put(-298,100){D4}
\put(-298,25){$\Delta x^6$}
\put(-298,-14){(a)}
\put(-215,125){NS5}
\put(-145,125){D6}
\put(-173,20){NS5}
\put(-175,-14){(b)}
\put(-25,-14){(c)}
\put(-137,5){$\Delta x^9$}
\put(-185,100){D4}
\put(-175,59){D4}
\put(-188,82){D2}
\put(-65,125){NS5}
\put(-87,98){NS5}
\put(-73,45){D2}
\put(-30,75){D6}
\put(-45,100){D4}
\caption{The brane configuration relevant for the Abelian Higgs model. 
(a) On the D4-brane suspended between parallel NS5-branes, 4
 dimensional ${\cal N}=2$ U(1) theory is realized. (b) The FI term is
 turned on, and the effective theory is the Abelian Higgs model. 
The dashed line ending on the D4-branes shows a D2-brane (on a
 ``flavor'' D6-brane) which is identified with a vortex string. (c) The
 D2-brane (dashed line) is in a Coulomb phase of its effective field
 theory. Because it can freely move away from the D4-brane, there is no
 solitonic interpretation in 3+1 dimensions. }
\label{ns5fig}
\end{center}
\end{minipage}
\end{center}
\end{figure}

The ${\cal N}=2$ Abelian Higgs model in 4 dimensions can be realized as
a theory on a D4-brane suspended between two parallel NS5-branes in Type
IIA superstring theory: see Fig.~\ref{ns5fig}(a). We have in addition
several D6-branes perpendicular to the other branes, to realize
hypermultiplets. Turning on the Fayet-Illiopoulos parameter $v_{\rm AH}$
corresponds to a translation of one of the NS5-branes along $x^9$. The
vortex strings are realized as D2-branes placed on the D6-branes and
suspended between the D4-branes, see Fig.~\ref{ns5fig}(b). These
D2-branes correspond to the vortex strings, as they possess the correct
charges and amount of supersymmetries. The bosonic part of the effective
field theory on two D2-branes is \cite{HT}  
\begin{eqnarray}
&& S_{\rm vortex}= \int\! dt dx \;
{\rm Tr}\left[
-\frac{1}{4g^2}F_{\mu\nu}F^{\mu\nu} 
-{\cal D}_\mu Z^\dagger {\cal D}^\mu Z 
-{\cal D}_\mu \psi^\dagger {\cal D}^\mu \psi
\right.
\nn\\
&& \hspace{70mm}
\left.
-\frac{g^2}{2}\left(
\psi\psi^\dagger - [Z,Z^\dagger]-r\unit_{2\times 2}
\right)^2
\right]
\label{lagvs}
\end{eqnarray}
which is a $1+1$ dimensional U(2) gauge theory with a complex adjoint
field $Z$ and a vector field $\psi$. Note that the theory becomes 2
dimensional since we took a limit where the dynamics along $x^9$ can be
ignored. We appropriately rescaled the adjoint scalar fields $\Phi^i$
and introduced a complex field representation 
$Z \propto \Phi^2 + i\Phi^3$.  
The Lagrangian includes terms of the D-string action (\ref{effe}), but
two features peculiar to (\ref{lagvs}) is that there are a FI parameter
$r$ and a new field $\psi$. This $r$ and the gauge coupling appearing in
the effective theory are related to the string theory parameters and the
original Abelian Higgs model parameters by
\begin{eqnarray}
 \frac{1}{g^2} = \frac{l_{\rm s} \Delta x^9}{g_{\rm s}}
= (2\pi)^3 l_{\rm s}^4v_{\rm AH}^2 \ , 
\qquad 
r = \frac{\Delta x^6}{2\pi g_{\rm s}l_{\rm s}}
 = \frac{2\pi}{g_{\rm AH}^2} \ .
\end{eqnarray}
Here $v_{\rm AH}$ and $g_{\rm AH}$ are the FI parameter and the gauge
coupling, respectively, in the original Abelian Higgs model.
As shown in \cite{HT}, we have to take the strong coupling limit
$g\rightarrow \infty$ so that the original Abelian Higgs model is
decoupled from other stringy modes. Therefore, the dynamics of the 
vortex strings is dictated solely by the moduli space of the theory
(\ref{lagvs}). This moduli space is defined by the D-term condition
\begin{eqnarray}
 \psi \psi^\dagger - [Z,Z^\dagger] - r \unit_{2\times 2} = 0 \ .
\label{D-term}
\end{eqnarray}
Thus, the effective theory of the vortex strings is a sigma
model whose target space is defined with this D-term equation.
It describes any slow motion of the vortex strings.\footnote{
Though we have used a specific brane realization of the Abelian Higgs
model, there are many other brane configurations. However, any embedding
of the model in string theory will reproduce the same result for the
effective theory on the vortex string, after taking the decoupling limit
(the low energy limit).} 

The important fact is that the D-term equation (\ref{D-term}) does not
allow the configuration (\ref{passing}) due to the existence of the FI
parameter $r$. The configuration (\ref{passing}) gives vanishing
commutator $[\Phi_2, \Phi_3]=0$ which cannot satisfy (\ref{D-term})
because the $2\times 2 $ matrix $\psi\psi^\dagger$ is of rank 1.
Thus, there is no naive classical solution of vortex
strings passing through each other.

If we take $r=0$, the Lagrangian (\ref{lagvs}) mostly reduces to
the D-strings action (\ref{effe}). In the brane configuration, this may
be achieved by putting $\Delta x^6=0$, then the D4- and D6-branes become
irrelevant for the motion of the D2-branes, and thus there are no
fundamental fields and no FI term in the D2-brane action.  
In this limit $\Delta x^6=0$, there is no solitonic interpretation of 
the D2-brane in the 4 dimensional theory. See Fig.~\ref{ns5fig}(c). 
The D2-branes (macroscopically $\sim$ D-strings) can pass through each
other classically, as the solution (\ref{passing}) shows. On the other
hand, when $r\neq 0$, the theory acquires the fundamental field $\psi$
and is in a Higgs phase, then the configuration (\ref{passing}) is not
allowed on the moduli space. This classical fact is the origin of the
difference between the reconnection property of colliding strings. 

In the next section, we introduce appropriate coordinates to
parametrize the moduli space (\ref{D-term}) and solve the sigma model
to show the inevitability of the reconnection of the colliding vortex
strings. In Section \ref{sec4}, we explain how a quantum mechanics on 
the classical D-string solution (\ref{passing}) 
will cause the reconnection.

\section{Reconnection of colliding vortex strings}
\label{sec3}

\subsection{Effective field theory on vortex strings}

The Moduli space metric on this Higgs branch defined by the D-term
condition (\ref{D-term}) was determined by a K\"ahler quotient technique 
in \cite{kim}, which we shall utilize in the following.
The D-term condition (\ref{D-term}) can
be solved by the parametrization \cite{kim}
\begin{eqnarray}
 Z = w \unit_{2 \times 2} + z
\left(
\begin{array}{cc}
1 & \sqrt{2b/a} \\ 0 & -1
\end{array}
\right), \quad 
 \psi = \sqrt{r}
\left(
\begin{array}{c}
\sqrt{1-b} \\ \sqrt{1+b}
\end{array}
\right)
\label{param}
\end{eqnarray}
where 
\begin{eqnarray}
 a \equiv \frac{2|z|^2}{r} \ , 
\quad b \equiv \frac{1}{a + \sqrt{1+a^2}} \ .
\end{eqnarray}
The parameter $w$ describes the center-of-mass for the two vortex
strings, while $2z$ parametrizes the relative position of them.
This can be seen when real or imaginary part of the matrix $Z$
is diagonalized. (These cannot be diagonalized simultaneously, though.)
Using this parametrization, the metric on the moduli space spanned by
$z$  is given by \cite{kim}
\begin{eqnarray}
 ds^2 = g(|z|) dz d\overline{z} \ , \quad 
g(|z|)\equiv \frac{|z|^2}{\sqrt{|z|^4 + r^2/4}}\ .
\end{eqnarray}
The center of mass position $w$ is decoupled from $z$ in this
approximation of slow motion on the moduli space, so we concentrate on
$z$. The effective action of the relative motion of the two
vortex strings is then written as 
\begin{eqnarray}
 S = {\cal T}\int \!dt dx \; g(|z|) \p_\mu z(t,x) 
\p^\mu \overline{z}(t,x) \ .
\label{effevs}
\end{eqnarray}
The approximation is valid if the lagrangian (the integrand in the above
$S$) itself is small
compared to 1. ${\cal T}$ is an analogue of the tension of 
strings, but this overall factor is irrelevant in the following
classical computations.

\begin{figure}[tp]
\begin{center}
\begin{center}
\includegraphics[width=15cm]{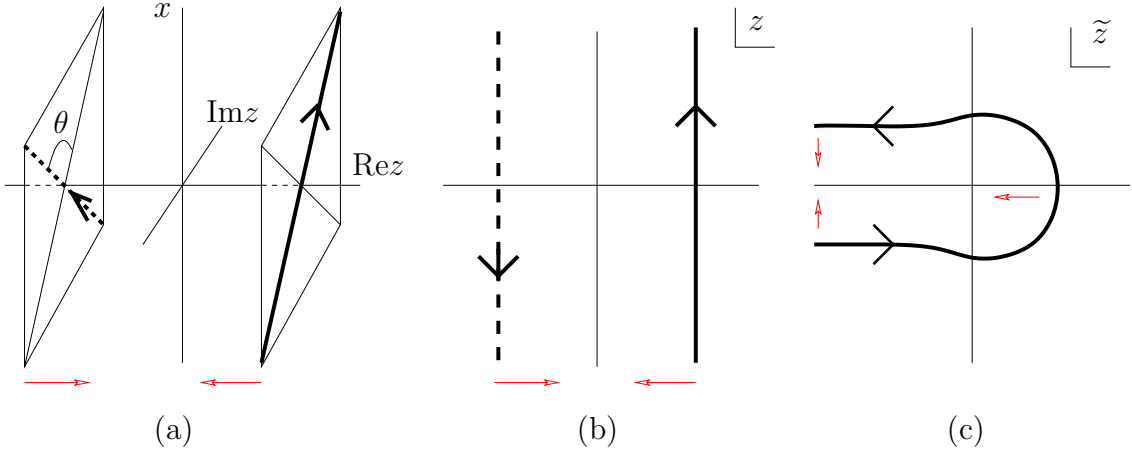}
\put(-370,-20){(a)}
\put(-370,140){$x$}
\put(-350,100){Im$z$}
\put(-408,96){$\theta$}
\put(-295,80){Re$z$}
\put(-145,135){$z$}
\put(-15,130){$\widetilde{z}$}
\put(-210,-20){(b)}
\put(-70,-20){(c)}
\end{center}
\begin{minipage}{14cm}
\begin{center}
\caption{(a) Initial motion of the two vortex strings (the thick line
 and the dashed thick line with arrows indicating their orientations). 
Small arrows show the direction of the motion. (b) Projection of (a)
 onto the plane spanned by the complex coordinate 
$z$. The dashed line is a mirror partner of
 the thick line. (c) Configuration of (a)(b) mapped onto the plane
 spanned by $\widetilde{z}$. } 
\label{inifig}
\end{center}
\end{minipage}
\end{center}
\end{figure}

The metric $g(|z|)$ describes a resolved cone geometry. This can be read
from the asymptotic behavior of the metric,
\begin{eqnarray}
 g(|z|) \sim \left\{
\begin{array}{ll}
2|z|^2/r & \mbox{for}\;\;\; |z| < \sqrt{r/2}\\
1 & \mbox{for}\;\;\; |z| > \sqrt{r/2} 
\end{array}
\right.
\end{eqnarray}
For large separation $|z|\gg \sqrt{r/2}$, the metric of course goes to
the flat metric. But when the vortex strings are close ($z\sim 0$), 
we can make a
coordinate transformation
\begin{eqnarray}
 \widetilde{z} \sim \frac{1}{\sqrt{2r}}z^2
\label{zz2}
\end{eqnarray}
resulting in a metric $ds^2 \sim |d\widetilde{z}|^2$. That is, near
the origin a well-behaving coordinate is $\widetilde{z}$ rather than
$z$. As is obvious from this expression of the coordinate
transformation,  the correct interpretation of $z$
is through the identification $z = \overline{z}$, that is, 
antipodal points in $z$ space are identified \cite{VS}. 
This fact is in agreement with the observation by Ruback \cite{Ruback}.

Now let us consider two colliding gauge strings. We consider an initial
condition at some time $t = t_{\rm ini}$ as 
\begin{eqnarray}
&& z = z_0 + i \tan(\theta/2) x  \\
&& \dot{z}= -\frac{v}{2}
\label{initialcond}
\end{eqnarray}
where the three parameters $\theta$, $z_0$ and $v$ are real and positive.
The configuration of two vortex-strings is shown in
Fig.~\ref{inifig}(a).  
Immediately one can deduce that 
$\theta$ is the relative angle of the strings, 
and $v$ is the relative velocity. Thus at $t=t_{\rm ini}$ the closest
distance between the strings is $2z_0$.
We assume that $v$ and $\theta$ are small so that we can use the 
moduli space approximation. This actual string configuration in the 3
dimensional space spanned by $x, z$ can be consistently projected onto
the 2 dimensional plane spanned only by $z$, see Fig.~\ref{inifig}(b). 
We may consider the relative motion of the strings on this projected
complex $z$-plane. 

\begin{figure}[tp]
\begin{center}
\includegraphics[width=6.5cm]{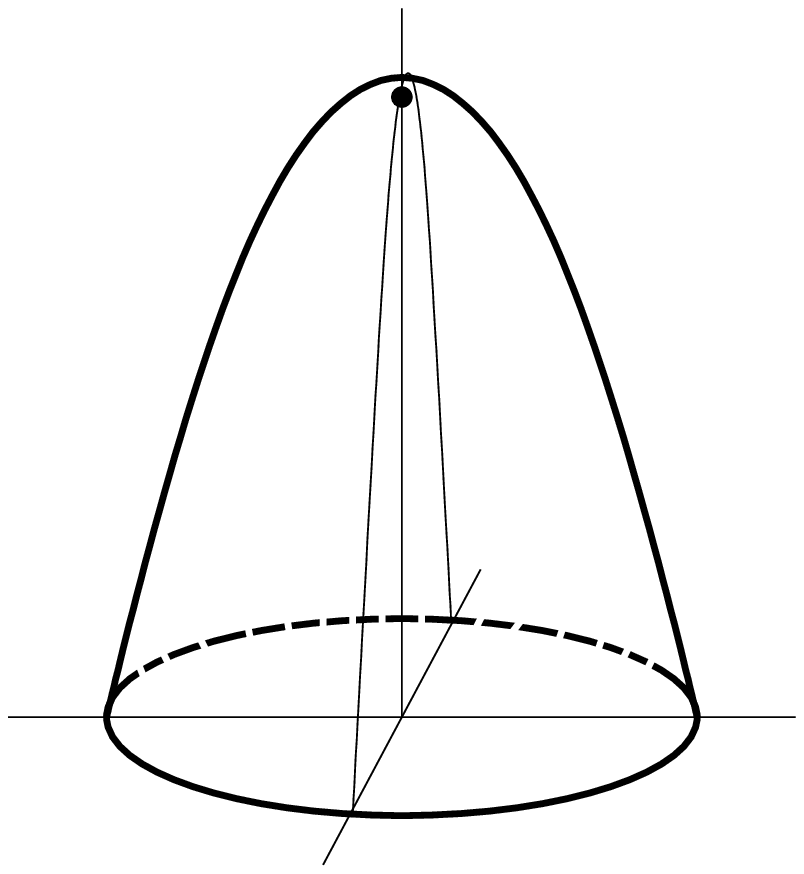}
\put(-105,195){$f$}
\put(-75,75){Im$\widetilde{z}$}
\put(0,25){Re$\widetilde{z}$}
\end{center}
\begin{center}
\begin{minipage}{14cm}
\caption{The moduli space metric is an induced metric on a 2 dimensional 
surface embedded in the 3 dimensional $f$-$\widetilde{z}$ space. This
 surface is a smoothed cone of a deficit angle $\pi$.
}
\label{cone1fig}
\end{minipage}
\end{center}
\end{figure}

Respecting the identification of the antipodal points, we make the
following coordinate transformation in the target space:
\begin{eqnarray}
 \widetilde{z} \equiv \frac{ z^2}{2(|z|^4+r^2/4)^{1/4}} \ .
\label{zz2p}
\end{eqnarray}
When $z\sim 0$, this transformation reproduces (\ref{zz2}).
This coordinate makes the geometrical picture of the metric clearer
as follows. The geodesic distance in this coordinate 
$\widetilde{z}\equiv \rho e^{i\varphi}$ is written as 
\begin{eqnarray}
 ds^2 = \left(1 + \left(\frac{d f(\rho)}{d\rho}\right)^2\right) 
d\rho^2 + \rho^2 d\varphi^2
\label{choi}
\end{eqnarray}
where $f(\rho)$ is a smooth function determined through (\ref{zz2p}).
We need only its asymptotic values,
\begin{eqnarray}
f(\rho) \sim -\sqrt{3}\rho \quad (\rho \sim \infty) \ , \quad
f(\rho) \sim -\sqrt{\frac{2}{r}} \rho^2 \quad (\rho\sim 0)\ .
\end{eqnarray}
The expression (\ref{choi}) shows that 
it describes a smeared surface of a cone with a deficit angle $\pi$. 
It is a metric induced on a 2 dimensional hypersurface in 3 dimensions
spanned by $\widetilde{z}=\rho e^{i\varphi}$ 
and a hypothetical new coordinate $f$,
with the embedding $f =f(\rho)$. The deficit angle $\pi$ can be read
from the asymptotic slope of $f$, 
$\lim_{\rho\rightarrow\infty}|df/d\rho|=\sqrt{3}$.
See Fig.~\ref{cone1fig}.

\begin{figure}[tp]
\begin{center}
\includegraphics[width=13cm]{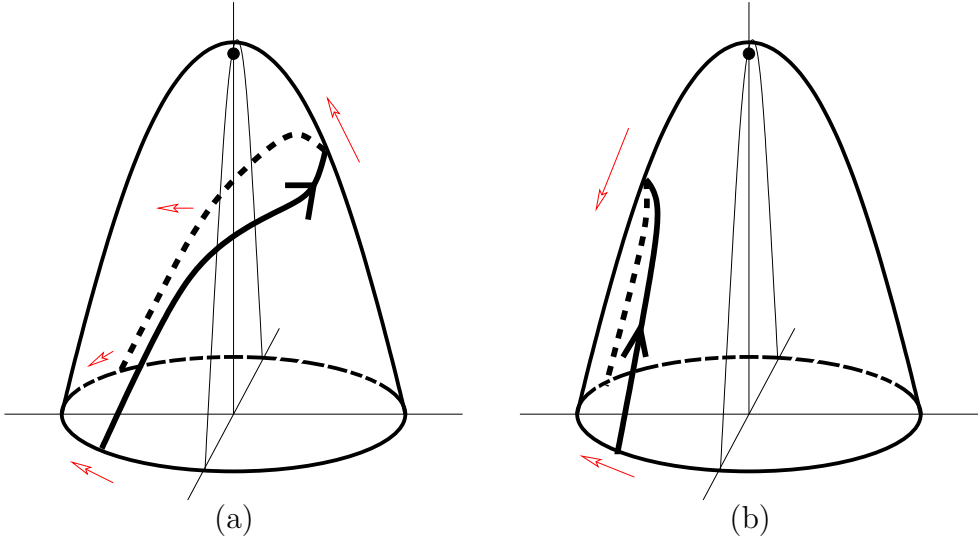}
\put(-290,-10){(a)}
\put(-95,-10){(b)}
\end{center}
\begin{center}
\begin{minipage}{14cm}
\caption{(a) The initial configuration of the Polyakov string (the thick
 line with an arrow) on the  cone corresponding to two straight vortex
 strings colliding. The small arrows show the
 direction of the motion of the Polyakov string. (b) The Polyakov string
 travels without feeling any singularity through the top of the cone,
 and arrives at the final configuration.   
}
\label{cone2fig}
\end{minipage}
\end{center}
\end{figure}

\subsection{Proof of reconnection of colliding vortex strings}

Note that for scattering of vortices (not the
vortex strings) in 2 spatial dimensions, the moduli space metric is the
same as (\ref{choi}). The initial condition of the colliding vortices is 
(\ref{initialcond}) with $\theta=0$: this is a simple dimensional
reduction along $x$ from the vortex string case. And it is obvious that
the worldline trajectory of the vortex is just a straight line
going through the top of the cone  --- a straight line in
the $\widetilde{z}$-space. In terms of $z$, this means a right-angle
scattering \cite{Ruback,VS}.

\begin{figure}[tp]
\begin{center}
\includegraphics[width=15cm]{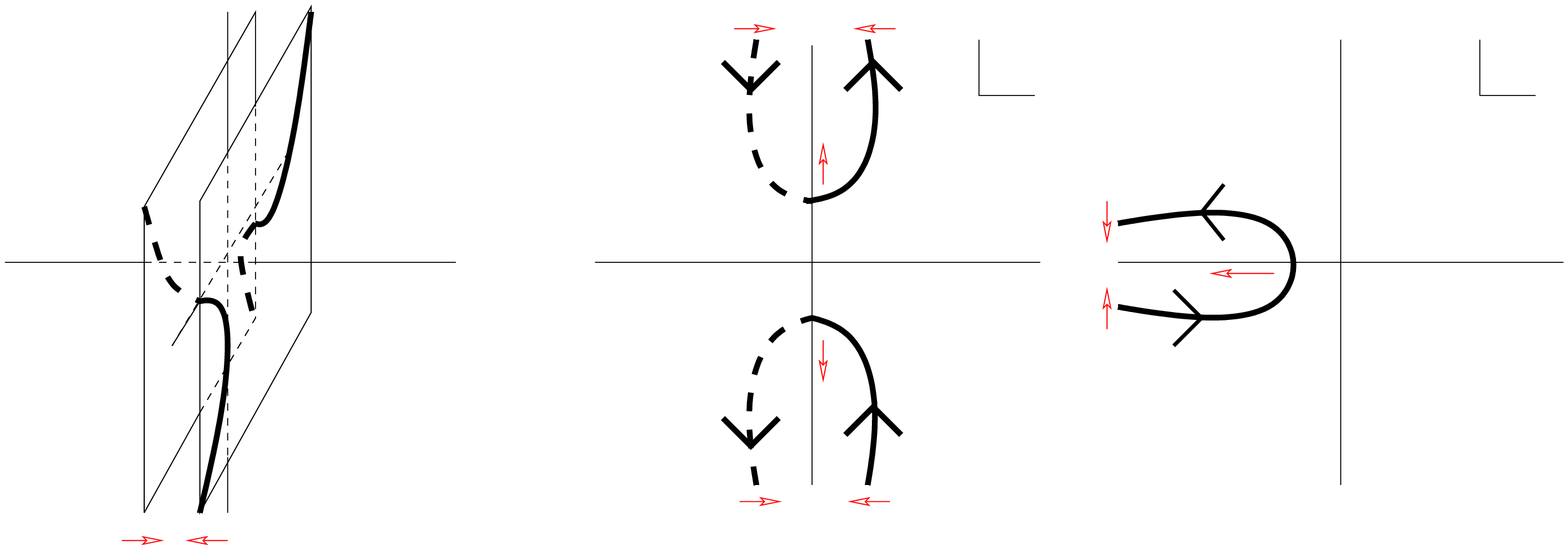}
\put(-370,-20){(a)}
\put(-375,140){$x$}
\put(-305,80){Re$z$}
\put(-155,130){$z$}
\put(-18,130){$\widetilde{z}$}
\put(-210,-20){(b)}
\put(-70,-20){(c)}
\end{center}
\begin{center}
\begin{minipage}{14cm}
\caption{(c) The string configuration in $\widetilde{z}$ space after a
 while. (b) The configuration of (c) is mapped to the space $z$. (a) It
 is lifted back to the original 3 dimensional space. The vortex strings
 are reconnected. The thick line is now connected with its original
 mirror string (dashed thick line).   }
\label{finalfig}
\end{minipage}
\end{center}
\end{figure}

The motion of the colliding vortex strings is described by a Polyakov
string whose motion is constrained on this smeared cone with a
particular initial motion. The initial condition (\ref{initialcond}) 
is given as Fig.~\ref{inifig}(c) in the $\widetilde{z}$-plane, and if we
map it onto the cone, it is a Polyakov string winding the top 
of the smeared cone, see Fig.~\ref{cone2fig}(a). The Polyakov string
moves slowly toward the top of the hill, and because the top is smeared
the string can smoothly travel beyond the top and come down, as in
Fig.~\ref{cone2fig}(b), which we call a final configuration. On the
projected $\widetilde{z}$-plane, the shape of the final configuration is
shown in Fig.~\ref{finalfig}(c). Mapping this back to the original 3
dimensional space, we obtain Fig.~\ref{finalfig}(a), which shows that 
the vortex strings are reconnected. We saw here that a smooth travel of
the Polyakov string on the smeared cone geometry turns out to give the
reconnection 
of the original colliding vortex strings. This proof is valid as long
as the sigma model description (\ref{effevs}) does not break down. Thus,
the reconnection always occurs for small collision velocity $v\ll 1$
and small intersection angle $\theta\ll 1$.

\begin{figure}[ht]
\begin{center}
\includegraphics[width=15cm]{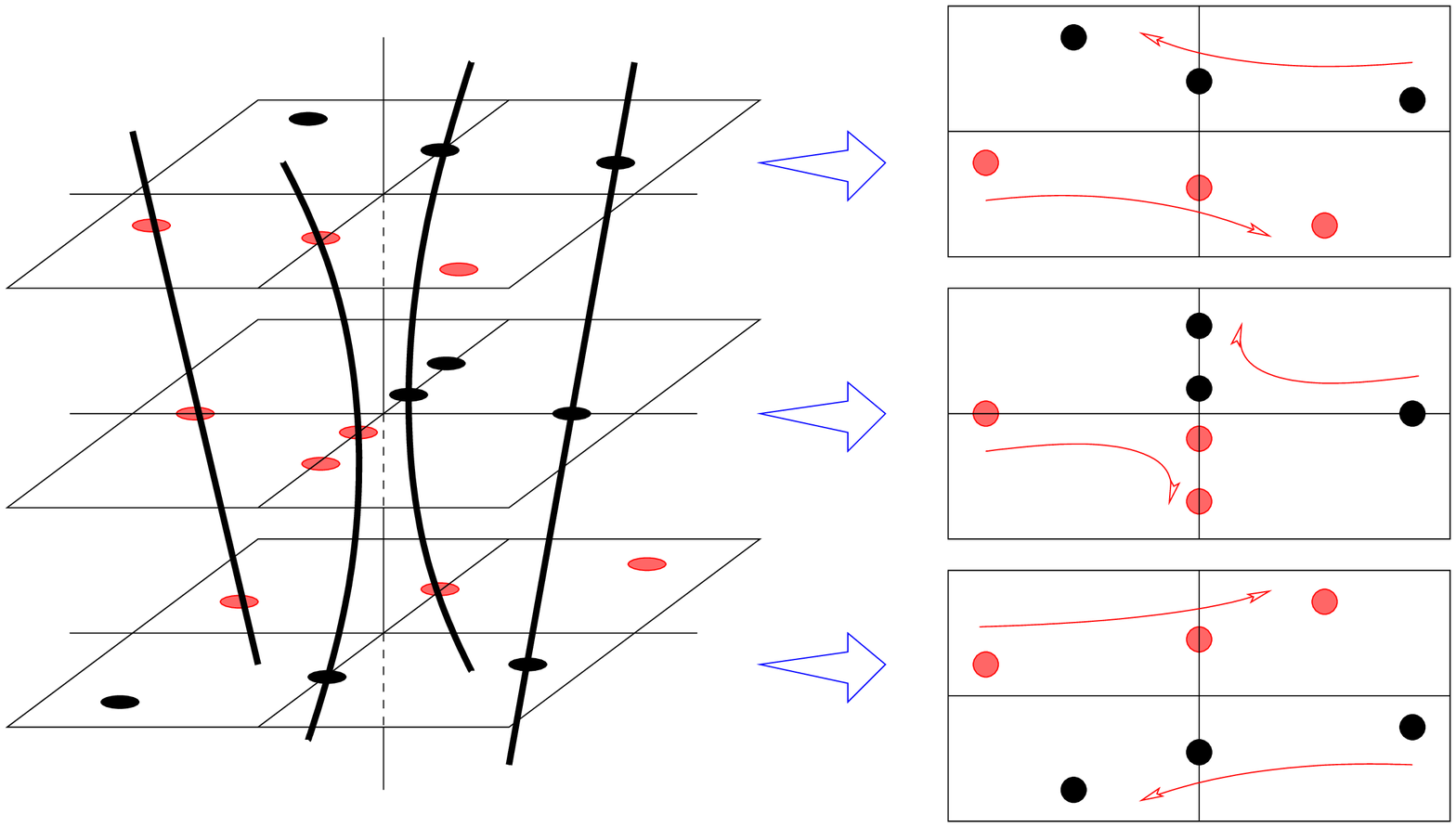}
\put(-310,230){$x$}
\put(-230,105){Re$z$}
\end{center}
\begin{center}
\begin{minipage}{14cm}
\caption{Reconnection of the colliding strings (Left) can be understood
 as a  collection of colliding vortices in 2 dimensions with various
 impact  parameters (Right). At the slice $x=0$, the vortex string
 reconnection is equivalent to the right-angle scattering of the
 vortices. }
\label{vortcollfig}
\end{minipage}
\end{center}
\end{figure}

If we take a slice $x=0$ of the vortex strings, it can be interpreted as
a scattering of a pair of vortices in 2 dimensions. 
For any slice of a fixed value of $x$, 
the reconnection of the colliding vortex strings can be seen as a
scattering of the vortices with various impact parameters given by 
(\ref{initialcond}) with the value of $x$ substituted. This is the
intuitive understanding of the reconnection, see Fig.~\ref{vortcollfig}.  
The slice $x=0$ has been considered first in Ref.~\cite{CT}.

\begin{figure}[tp]
\begin{center}
\includegraphics[width=13cm]{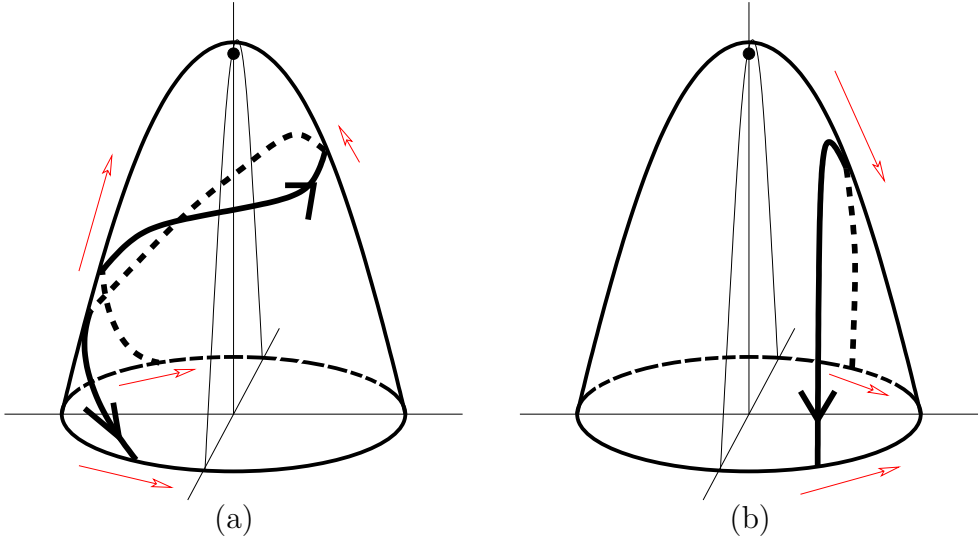}
\put(-290,-10){(a)}
\put(-95,-10){(b)}
\end{center}
\begin{center}
\begin{minipage}{14cm}
\caption{(a) When the relative velocity is large, the encircling 
speed of the  Polyakov string far from the tip of the cone (the speed
 at the bottom in the figure) is so large that the string 
self-intersects.
(b) As a consequence, after a while 
the Polyakov string is brought on the side of
 positive Re$\widetilde{z}$, but the orientation is opposite to the
 original configuration in Fig.5(a).
}
\label{velconefig}
\end{minipage}
\end{center}
\end{figure}

When the collision velocity is large enough, numerically the velocity
upper bound has been observed for the reconnection to occur
\cite{Shellard}.  
In this situation, the sigma model approximation becomes incorrect, 
but in the next subsection we derive the upper bound analytically 
by looking at a relativistic consistency condition for the shape of the
reconnected vortex strings. But before closing
this subsection, we would
like to give an intuitive picture of the existence of the upper bound,
using the figure of the cone, Fig.~\ref{cone1fig}, \ref{cone2fig}.
Suppose that even for vortex strings moving relativistically fast,
the cone geometry captures a correct dynamics.
When the relative velocity is too fast, the motion of the Polyakov
string at the bottom of the cone in the figures is very fast, and it
goes around the cone faster than the string passes the top of the
cone. See Fig.~\ref{velconefig}(a). Then the final configuration is not
in the region ${\rm Re}\;\widetilde{z}<0$ as in Fig.~\ref{cone2fig}(b) 
but comes back to the region ${\rm Re}\; \widetilde{z}>0$ 
(Fig.~\ref{velconefig}(b)). A careful look at the string reveals that
the orientation of the final Polyakov string is opposite compared to the
initial configuration of Fig.~\ref{cone2fig}(a). This means that in the
original spacetime the vortex strings pass through each other without
reconnection. 

\subsection{Static vortex strings: reconnection via pair annihilation}

We have seen that the map to the Polyakov string on the cone geometry
provides a physical and clear understanding of the reconnection of the
colliding vortex strings. It would be better to confirm the dynamics
by explicitly solving the equations of motion of  the sigma
model action (\ref{choi}) with the given initial condition
(\ref{initialcond}), but it 
turns out to be complicated. In this subsection we solve them with a
static ansatz, which provides an evidence supporting the inevitability
of the reconnection for vortex strings. The static ansatz can be
understood as an adiabatic collision of the vortex strings, that is, an
infinitesimally small collision velocity $v$. 
We will see that static configurations of a pair of the vortex strings 
tend to be aligned in orientations opposite to each other: locally
around the intersection point 
it becomes a string anti-string pair. Field theoretically the pair of
the vortex string and the anti vortex string should be unstable and
would decay to the vacuum (only around the intersection), 
then the resulting configuration should be
the reconnected vortex strings. Our result is coincident with
\cite{yaacov} in which a different effective analysis on the vortex
strings has been performed. 

The equations of motion for the field $z(t,x)\equiv z_1 + i z_2$ of the
sigma model (\ref{effevs}) in two dimensions are 
\begin{eqnarray}
 \p_\mu \left(g(|z|) \p^\mu z\right)=0 \ .
\label{eom}
\end{eqnarray}
The static ansatz $\p_0 z =0$ can be consistently imposed. Furthermore,
we restrict our attention to the case where ${\rm Re}z = z_1$ is a
real constant $z_0$.  
This means that one vortex string lies on a two dimensional
plane Re$z=z_0$ while the other lies on Re$z=-z_0$. Because $\p z_1=0$, 
this restriction is again consistent with the equations of motion,
thus the dynamical variable to be solved is only $z_2(x)$. 
The equation of motion in this case can be integrated immediately to
give
\begin{eqnarray}
 x = C \int_0^{z_2} \!\!\! g\; dy 
\end{eqnarray}
where the argument of $g$ is given by $|z|=\sqrt{z_0^2 + y^2}$,
and thus
\begin{eqnarray}
 g = \frac{z_0^2 + y^2}{\sqrt{(z_0^2 + y^2)^2 + r^2/4}} \ .
\end{eqnarray}
We have fixed an integration constant by putting $z_2(x=0)=0$
without losing generality, and $C$ is
another integration constant which can be determined as follows. 
The relative angle between the vortex strings can be seen in their
asymptotic slopes. Noting that above $g$ goes to the unity for large
$y$, we find $x \sim C z_2$ for large $z_2$. Therefore
\begin{eqnarray}
 C = 1/\tan (\theta/2) \ .
\end{eqnarray}

\begin{figure}[tp]
\begin{center}
\includegraphics[width=4cm]{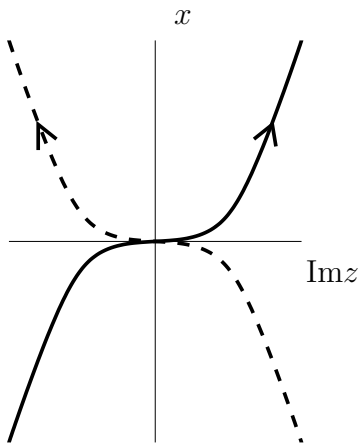}
\put(-0,62){Im$z$}
\put(-50,160){$x$}
\end{center}
\begin{center}
\begin{minipage}{14cm}
\caption{A static solution of two vortex strings lying on the
 same plane Re$z=0$. The dashed line is the mirror string (specified by
 $-z$). At the origin, the orientation of the solid line is opposite to
 that of the dashed line, showing the tendency to form a pair of string
 and anti-string. }
\label{staticfig}
\end{minipage}
\end{center}
\end{figure}

Because the function $g$ is a monotonically increasing
function of $z_2$ for fixed $z_1=z_0$, we deduce that $\p x / \p z_2$
is also monotonically increasing and approaching $1/\tan (\theta/2)$
asymptotically. The value of  $\p x / \p z_2$ at the origin $x=z_2=0$
is 
\begin{eqnarray}
\left. \frac{\p x}{\p z_2}\right|_{x=z_2=0}= \frac{1}{\tan(\theta/2)}
\frac{z_0^2}{\sqrt{z_0^4 + r^2/4}} \ .
\end{eqnarray}
The important point is that this vanishes for $z_0=0$. See
Fig.~\ref{staticfig}. For the value $z_0=0$, two vortex strings are on
the same two dimensional plane $z_1=0$. At the origin $z_1 = z_2 = x=0$
they intersect, and in fact become anti-parallel to each other: 
a pair of a vortex string and an anti vortex string. They should
annihilate with each other, and the resulting configuration should be
the reconnected vortex strings. The reasoning that the local pair
annihilation of strings leads to the reconnection is the same as what
was found in \cite{HN} in D-brane reconnection. 

The fact that two static vortex strings tend to intersect with 
relatively opposite orientation is a supporting evidence for the
inevitability of the reconnection of the colliding strings. Our result
is consistent with Ref.~\cite{yaacov} in which a different effective
description of the vortex strings was adopted.

\subsection{Velocity upper bound for reconnection of vortex strings}

The numerical simulations \cite{Shellard} 
indicate that there
is an upper bound for the relative velocity, for the vortex strings
to be reconnected. The ``probability'' of reconnection, termed in
\cite{Shellard} should be understood as a condition on the velocity
for the reconnection to occur classically. For vortex strings, the
reconnection is a purely classical phenomenon. 
There is a paper \cite{CT} which tried to derive in field theories 
this upper bound by analytic calculations concerning  
deformation of classical solutions of two coincident vortices with some
ansatz. The result is the following velocity upper bound,
\begin{eqnarray}
 \bar{v} 
< \sqrt{\frac{4\alpha (1-\cos\theta)}{1 + 4\alpha(1-\cos\theta)}} \ ,
\label{alpha}
\end{eqnarray}
where $\bar{v}$ is the velocity of the colliding vortex strings in the
center-of-mass frame (and thus related to the relative velocity $v$ 
as $v = 2\bar{v}/(1 + \bar{v}^2)$), 
and $\alpha$ is an unknown parameter introduced in the ansatz
in \cite{CT}. This dependence on $\theta$ 
coincides with the numerical results in \cite{Shellard} qualitatively. 
In this section we derive this velocity upper bound (\ref{alpha})
by a geometrical consideration, without referring to any classical field 
configuration. Thus, although the condition (\ref{alpha}) has been
derived with a specific field configuration in a certain theory, we show
that the same upper bound holds true for general vortex strings.

\begin{figure}[tp]
\begin{center}
\includegraphics[width=15cm]{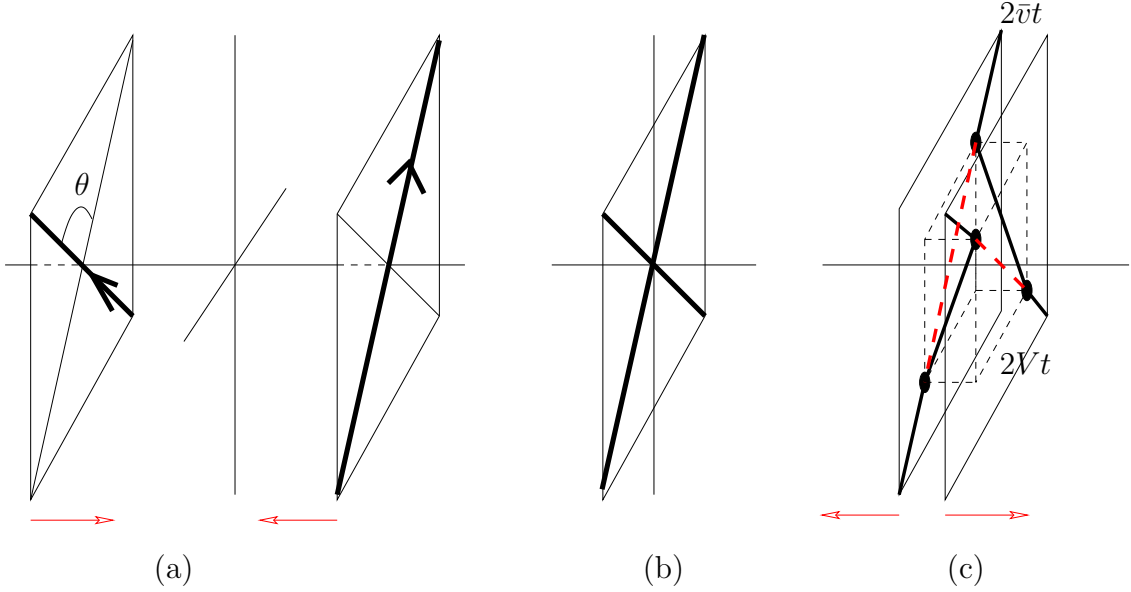}
\put(-50,57){$2Vt$}
\put(-50,190){$2\bar{v}t$}
\put(-370,-20){(a)}
\put(-400,125){$\theta$}
\put(-185,-20){(b)}
\put(-70,-20){(c)}
\end{center}
\begin{center}
\begin{minipage}{14cm}
\caption{Schematic figure of reconnection of vortex strings. (a) Two
 vortex strings (thick lines) are colliding. Small arrows indicate the
 directions of the string motion. (b) They collide at
 $t=0$. (c) They reconnect with each other. The shape after the
 reconnection is assumed in such a way that the reconnected part is 
on a dashed box which expands. The blobs represent kink points. The
 thick dashed lines are the strings which would have been present if
 the reconnection didn't occur.   } 
\label{reconne2}
\end{minipage}
\end{center}
\end{figure}

First we assume that the shape of the reconnected vortex strings is as
in Fig.~\ref{reconne2}, 
and neglect the effect of the energy of the kink points. 
In the center-of-mass frame, the velocity of the original strings is 
$\bar{v}$
and the velocity of the reconnected region is $V$. Generically $V$ may
depend on time but here we consider the configuration at times 
just a little after the collision
and so $t$ is close enough to the collision incidence $t=0$ 
and thus we may regard $V$ as a
constant. Note that the direction of the motion $V$ is perpendicular to
that of $\bar{v}$, due to our assumption on the shape of the strings. 

Let us consider the energy gain by assuming that the reconnection has
occurred,
\begin{eqnarray}
 \delta E = E_+-E_- \ .
\end{eqnarray}
Here $E_+$ is the energy produced by the reconnection, that is, the
energy of the string between the generated kinks. (In
Fig.~\ref{reconne2}, this is the energy of the solid lines along the box 
surface.) $E_-$ is the energy
of the original strings which disappeared after the reconnection.
(In Fig.~\ref{reconne2}, this is the energy of the dashed lines.)
Since the location of the kink points are
\begin{eqnarray}
 (Vt \cot(\theta/2), Vt, \bar{v}t) \ , \quad 
 (-Vt \cot(\theta/2), Vt, -\bar{v}t) \ ,
\label{kinkloc}
\end{eqnarray}
an explicit calculation gives
\begin{eqnarray}
 E_+ = 4 {\cal T} \frac{t\sqrt{\bar{v}^2 
+ V^2 \cot^2(\theta/2)}}{\sqrt{1-V^2}}
\ . 
\end{eqnarray}
The denominator comes from the gamma factor for relativistic motion of
the strings. The numerator is the length of the reconnected strings.
On the other hand, the loss of the energy $E_-$ is 
\begin{eqnarray}
 E_- = 4 {\cal T} \frac{Vt }{\sqrt{1-\bar{v}^2}\sin(\theta/2)} \ .
\end{eqnarray}
Thus we obtain the total energy gain,
\begin{eqnarray}
\delta E = 4 {\cal T}t \left(
\frac{\sqrt{\bar{v}^2 + V^2 \cot^2(\theta/2)}}{\sqrt{1-V^2}}
-\frac{V }{\sqrt{1-\bar{v}^2}\sin(\theta/2)} 
\right) \ .
\end{eqnarray}
Given an initial velocity $\bar{v}$, if this energy gain can be negative
for an appropriately chosen velocity $V$, then the reconnection should
occur.  

\begin{figure}[tp]
\begin{center}
\includegraphics[width=12cm]{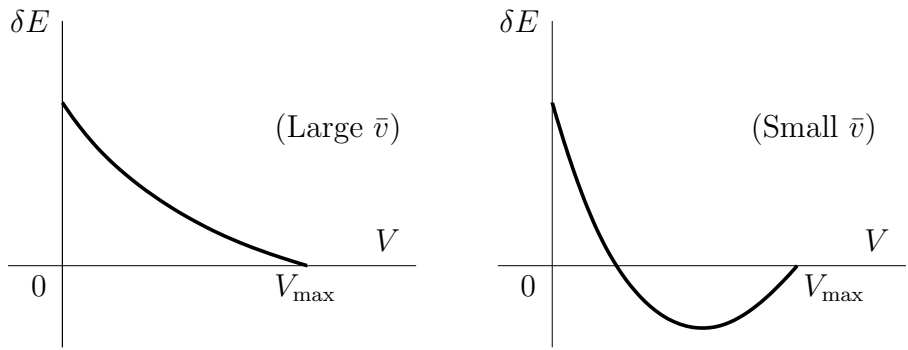}
\put(-340,120){$\delta E$}
\put(-240,20){$V_{\rm max}$}
\put(-240,80){(Large $\bar{v}$)}
\put(-332,20){$0$}
\put(-202,37){$V$}
\put(-155,120){$\delta E$}
\put(-40,20){$V_{\rm max}$}
\put(-60,80){(Small $\bar{v}$)}
\put(-147,20){$0$}
\put(-17,37){$V$}
\end{center}
\begin{center}
\begin{minipage}{14cm}
\caption{The energy gain in the allowed region of $V$. For large
 $\bar{v}$, there is no value of $V$ in the allowed region 
 $0<V<V_{\rm max}$ which gives a negative $\delta E$ (Left), while for
 small $\bar{v}$ there exist such values (Right). } 
\label{graphfig}
\end{minipage}
\end{center}
\end{figure}

We note here that a-priori there is a restriction on the
value of $V$, because the kink points cannot propagate faster than the
speed of light. 
This condition can be easily read from the kink coordinates
(\ref{kinkloc}), giving a constraint
\begin{eqnarray}
 0\leq V \leq \sqrt{1-\bar{v}^2}\sin(\theta/2)\; (\equiv V_{\rm max}) \ .
\label{allowed}
\end{eqnarray}
At the allowed velocity minimum $V=0$, the energy gain $\delta E$ is
positive, while at the allowed velocity maximum
$V = V_{\rm max}$, the energy gain vanishes (see Fig.~\ref{graphfig}).
Checking the sign of the derivative $\p (\delta E) / \p V$, 
we find the condition for $\bar{v}$ to provide a negative $\delta E$
with a suitably chosen $V$ in the allowed region (\ref{allowed}), as 
\begin{eqnarray}
 \bar{v} < \frac{\sin(\theta/2)}{\sqrt{1 + \sin^2(\theta/2)}} \ .
\label{upper}
\end{eqnarray} 
Thus we have an upper bound for $\bar{v}$ to have the reconnection. 
This bound is derived only from a geometrical consideration, but 
surprisingly, it is the same as (\ref{alpha}) found in
\cite{CT}, if we choose the parameter $\alpha=1/8$.\footnote{ 
In the present evaluation of the velocity upper bound, we assumed the
shape of the reconnected vortex strings and the position of the kinks. 
Thus it is possible that the true upper bound may be a little bigger
than what we have obtained here, (\ref{upper}).}

\section{Reconnection of colliding D-strings}
\label{sec4}
\setcounter{footnote}{0}

In this section we give an estimation of the probability of
reconnection of two D-strings when they collide with a constant relative
velocity, by using a low energy effective field theory on D-strings
(\ref{effe}). As shown in Section \ref{sec2}, 
there is a classical solution
of colliding D-strings passing through each other without
reconnection. The reconnection is triggered by {\it quantum tachyonic 
fluctuations} around the classical solution, as we will see.
Thus the crucial point is that the reconnection is actually a 
probabilistic
event, in contrast to the case of the classical reconnection of the
colliding vortex strings studied in Section \ref{sec3}.
In the following, we compute the reconnection probability of the
colliding D-strings whose classical trajectory is given by
(\ref{passing}). 
We shall compare our result 
with the calculation by the string worldsheet theory 
given in \cite{JJP} and find a
qualitative agreement.

\subsection{Tachyon condensation and reconnection}

First let us briefly review how the reconnection of intersecting
D-strings is realized by the tachyon condensation in terms of Yang-Mills
theory on the D-strings \cite{HN}.
For intersecting D-branes, there appear tachyonic fluctuations around
the intersection point, and the condensation of them shows a
reconnection. This has been explicitly demonstrated in \cite{HN}, and
further studied in \cite{TH1, TH2, Na}.\footnote{For related earlier
references, see \cite{Aki-Wati, Morosov}.} 

We start with the D-string action (\ref{effe}) which is 
the $1+1$ dimensional $SU(2)$ 
Yang-Mills theory. Precisely, this is a low-energy effective action of
two coincident parallel D-strings. Following \cite{HN}, we turn on an
intersection angle $\theta$ (while keeping the D-strings on a 
two-dimensional plane, $\bar{v}=0$ in (\ref{passing}) and thus 
$\Phi_3=0$) and perform a fluctuation analysis, to get the lowest
fluctuation mode 
\begin{eqnarray}
 \Phi_2 = \frac{T(t)}2 
\left(
\begin{array}{cc}
0 & 1 \\ 1 & 0
\end{array}
\right)
\exp\left[
-\frac{\tan (\theta/2)}{2\pi l_{\rm s}^2}x^2
\right] \, , 
\\
 A_x = \frac{T(t)}2 
\left(
\begin{array}{cc}
0 & -i \\ i & 0
\end{array}
\right)
\exp\left[
-\frac{\tan (\theta/2)}{2\pi l_{\rm s}^2}x^2
\right] \, .
\label{phi1}
\end{eqnarray}
Here $x$ is the coordinate along the original D-string worldvolume,
and we took the gauge $A_0=0$. 
This fluctuation eigen-mode has the mass
squared 
\begin{eqnarray}
 m^2 = -\frac{\tan(\theta/2)}{\pi l_{\rm s}^2} 
+ \frac{(2z_0)^2}{(2\pi l_{\rm s}^2)^2}
\label{tacmas}
\end{eqnarray}
with $z_0=0$. The parameter $z_0$ appearing here becomes non-zero
if another transverse separation along $\Phi_3$ between the D-strings
is turned on as $2\pi l_{\rm s}^2\Phi_3 = {\rm diag} (z_0, -z_0)$.
(This $z_0$  has been set to zero in \cite{HN}.)
Inclusion of nonzero constant $z_0$ in the fluctuation analysis 
is straightforward, and the result (\ref{tacmas}) coincides with the 
string worldsheet analysis \cite{BDL} for $\theta \ll 1$. 
In addition, there are infinite number of other eigen-modes with higher
mass squared, which reproduce a part of the worldsheet spectra of a
string connecting the two D-strings \cite{HN, TH2, Na2, epple}.

Knowing the eigen-mode decomposition along the direction $x$, we can
reduce the system by one dimension lower, by integrating over $x$,
to obtain an effective action for the coefficient function $T(t)$ 
in (\ref{phi1}).
This coefficient function is the tachyon field, because when the
separation $2z_0$ is small enough, the mass squared (\ref{tacmas})
becomes negative.
Substituting the profile (\ref{phi1}) 
back to the Yang-Mills action (\ref{effe}), we obtain
\begin{eqnarray}
 S &= &\frac12{\cal T}_{\rm D1} (2\pi l_{\rm s}^2)^2\int dt dx 
\left[ (\p_t T(t))^2 - m^2 T^2 \right] 
\exp\left[
-\frac{2\tan (\theta/2)}{2\pi l_{\rm s}^2}x^2
\right] \nn\\
& =& \frac{1}{g_{T}}\int dt 
\left[ \frac12(\p_t T(t))^2 - \frac12m^2 T^2 \right] \ .
\label{tacac}
\end{eqnarray}
This is the tachyon action in one dimension which we use in this
section. The overall coefficient is related to the original parameters
as 
\begin{eqnarray}
 g_T = \frac{g_s \sqrt{\theta}}{2\sqrt{2}\pi^2 l_s^3} \ .
\end{eqnarray}

We may regard the action (\ref{tacac}) 
as a quantum mechanics of a particle moving in a harmonic oscillator:
in this interpretation, the value $T$ is the position of a particle, and
the overall factor $1/g_T$ is the mass of the particle. 

When the tachyon acquires a vev, this immediately leads to the
reconnection. The eigen-mode $T$ enters in the off-diagonal entries of
the matrix $\Phi$ as in (\ref{phi1}) and thus,
diagonalizing the matrix Higgs field $\Phi_2$, we obtain the transverse
displacement of the D-strings (for $z=0$) as
\begin{eqnarray}
 2\pi l_{\rm s}^2 \phi = \pm 
\sqrt{(\tan(\theta/2))^2 x^2 + (\pi l_{\rm s}^2)^2 \exp
\left[-\frac{\tan (\theta/2)}{\pi l_{\rm s}^2}x^2\right] T^2} \ .
\end{eqnarray}
Therefore, the tachyon vev is related to the separation of the
reconnected D-strings $\Delta$ (the closest points on the D-strings 
are given by $x=0$) as 
\begin{eqnarray}
 \Delta = 2\pi l_{\rm s}^2 T \ .
\label{deltae}
\end{eqnarray}
We assume that this relation holds also for $z\neq 0$ in the following
analysis of D-string collision.\footnote{For non-zero $z_0$, 
the two Higgs fields $\Phi_2$ and $\Phi_3$ are not simultaneously
diagonalizable and thus this geometrical interpretation is not
rigorous.} 

\subsection{Time evolution of tachyon wave function and reconnection
  condition}

The tachyon mass squared $m^2$ is negative when the D-strings are close
enough to each other, but otherwise, it is positive and gives a 
usual harmonic oscillator. Thus the collision of the angled D-strings 
provides time-dependent transitions between these two situations.
To describe the collision of the angled D-strings, we turn on a small
relative velocity $v(>0)$,  
\begin{eqnarray}
 2z_0 = v t \ ,
\end{eqnarray}
as in (\ref{passing}).
For very small velocity, the fluctuation analysis in the previous
subsection is valid ($v \ll \theta \ll  1$).
The time when the tachyon mass squared becomes zero is 
\begin{eqnarray}
t = \pm t_0 \ , \quad  t_0 = \frac{l_s\sqrt{2\pi\theta}}{v} \ .
\label{tzero}
\end{eqnarray}
The one-dimensional action (\ref{tacac})
is quite simple, a harmonic oscillator with
a time-dependent frequency. 
The frequency becomes imaginary for the period
\begin{eqnarray}
 -t_0 < t <  t_0
\end{eqnarray}
during which the harmonic potential becomes upside down and 
there exists a tachyonic instability.

\begin{figure}[tp]
\begin{center}
\includegraphics[width=15cm]{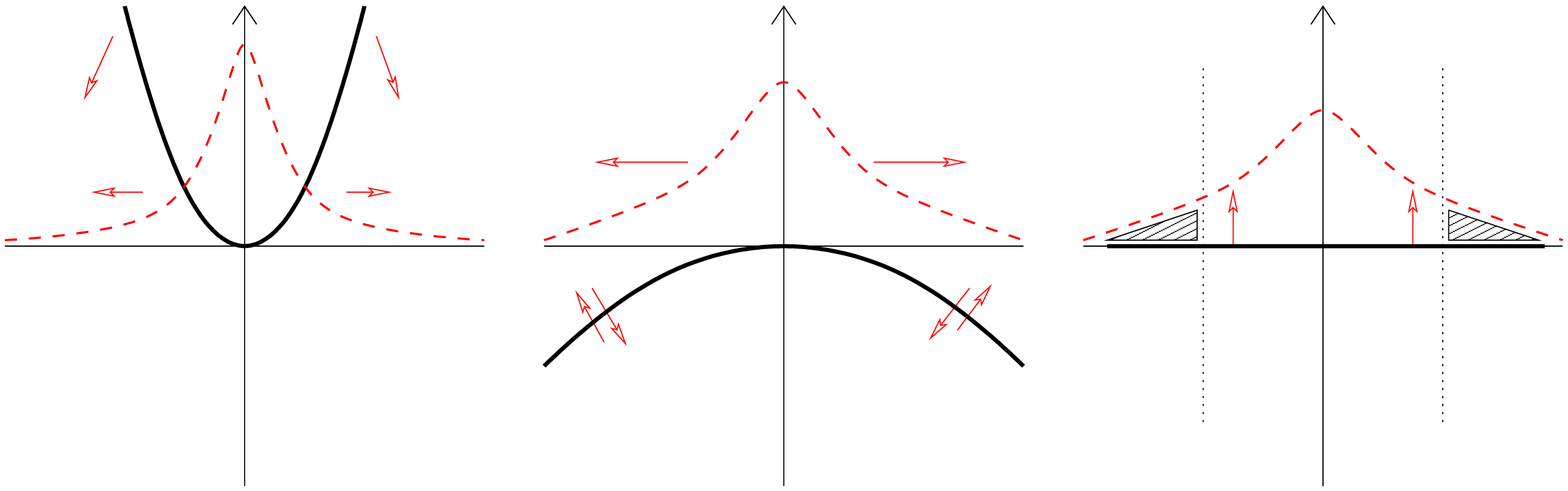}
\put(-370,-20){(a)}
\put(-375,140){$\psi(T,t)$, $V(T)$}
\put(-305,50){$T$}
\put(-220,-20){(b)}
\put(-70,-20){(c)}
\end{center}
\begin{center}
\begin{minipage}{14cm}
\caption{Time evolution of the harmonic potential (thick lines) and the
 tachyon wave function (dashed lines). The small arrows denote their
 time evolution. (a) When the D-strings are far
 enough from each other ($t<-t_0$), 
the tachyon mass squared is still positive and
 the wave function is a gaussian. (b) For $-t_0<t<t_0$, the potential
 becomes inverse-harmonic, and the wave function expands rapidly. (c) At 
 $t=t_0$, the system gets back to the bounded harmonic potential. For 
 large tachyon values the reconnection will occur. Thus the reconnection
 probability is evaluated by integrating the shaded region
 of the wave function. }
\label{tfig}
\end{minipage}
\end{center}
\end{figure}

\begin{figure}[tp]
\begin{center}
\includegraphics[width=15cm]{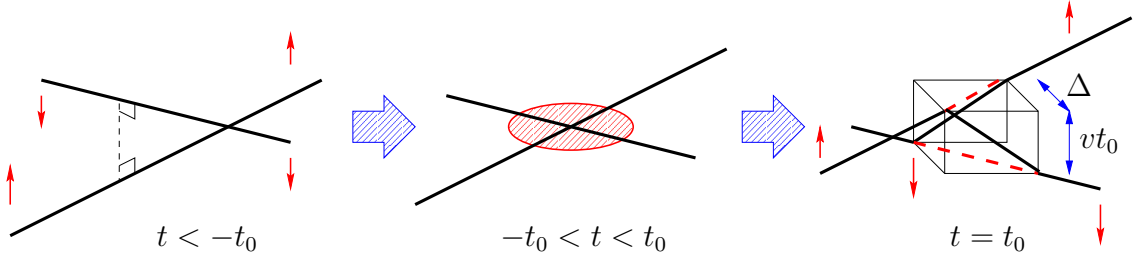}
\put(-26,57){$\Delta$}
\put(-21,37){$vt_0$}
\put(-370,0){$t<-t_0$}
\put(-240,0){$-t_0<t<t_0$}
\put(-70,0){$t=t_0$}
\end{center}
\begin{center}
\begin{minipage}{14cm}
\caption{Three stages for the reconnection of the D-strings. Two
 D-strings are approaching (Left), and tachyon wave function spreads
 (Center). At last, the D-strings are reconnected (Right). The dashed
 lines indicate the case where D-strings were not
 reconnected. Essentially this figure is the same as Fig.~10, 
but just
 pushed sideways to show the reconnected region explicitly.
}
\label{reconne}
\end{minipage}
\end{center}
\end{figure}

The physical picture of quantum evolution of the system is as
follows.\footnote{A quantum analysis of the 
reconnection was studied in \cite{sato}.} 
For large 
negative $t$, the system is with the harmonic potential and
so the wave function $T(t)$
is a gaussian function whose width is determined by
the frequency changing slowly in time. Then at $t=-t_0$, the system
drastically changes to an unbounded one and the wave function begins to 
spread out very quickly. However, when $t=t_0$, the potential comes 
back to the original harmonic form and the system recovers to be a
bounded well-defined system. This means that the wave function rapidly 
spreads
out with the unbounded potential only for the finite period, 
$-t_0 < 0 < t_0$. See Fig.~\ref{tfig}.
At the end of the unstable era, $t=t_0$, we have a well-spread wave
function which allows a large value of $T(t)$ and gives the
reconnection. 
Then how can we know if the reconnection occurs or not? We introduce the
following geometrical requirement for the tachyon field to be observed
to give the reconnection. At the instant $t=t_0$, two D-strings will be
separated (asymptotically) by $vt_0$ in the $\Phi_3$ direction. Thus, as
seen in Fig.~\ref{reconne}, the reconnected distance $\Delta$ along
$\Phi_2$ should be larger than this $v t_0$ so that the reconnection
provides energy reduction. The geometrical requirement for the
reconnection to be observed at $t=t_0$ should be
\begin{eqnarray}
v t_0 < \Delta \ .
\end{eqnarray}
This is translated to the condition for the tachyon field, using
(\ref{deltae}) and (\ref{tzero}) as 
\begin{eqnarray}
 T > \frac{\sqrt{\theta}}{\sqrt{2\pi}l_s} \ .
\label{taccondi}
\end{eqnarray}
Therefore, to compute the reconnection probability for the colliding
D-strings, first we compute the time evolution of the tachyon wave
function, and at $t=t_0$ we evaluate the probability to have $T$
satisfying the condition (\ref{taccondi}). We will perform this in the
next subsection explicitly.

\subsection{Reconnection probability}

Let us consider the evolution of the tachyon wave function 
$\psi(T,t)$. Since it is technically difficult to get the exact
expression of this wave function for the time-dependent frequency, 
we adopt the following crude approximation for the time-dependence of
the frequency: 
\begin{eqnarray}
 m^2 = \frac{(vt)^2}{(2\pi l_s^2)^2} & (t<-t_0) 
\label{firstp}\\
 m^2 = -m_{\rm second}^2 \equiv 
-s\frac{\theta}{2\pi l_s^2} & (-t_0<t<t_0)
\label{secondp}
\end{eqnarray}
In the first period, the mass squared is positive and  proportional
to $t^2$, where we neglected the first term in (\ref{tacmas}).
This approximation is valid if $t \ll -t_0$, but we are going to use
this $m^2$ until $t=-t_0$. In the second tachyonic 
period, we approximate the mass
squared by its typical tachyonic value, and $s$ is an ${\cal O}(1)$
parameter. One may choose $s=2/3$ which is the average tachyon mass
squared in this second period. This approximation of the second period
is especially useful since we can apply the results obtained in
\cite{GP}, where time evolution of a gaussian wave function in the
tachyonic harmonic potential was calculated. 

It is easy to show that the technique developed in \cite{GP} can be
used also for non-constant mass squared such as the one in the first
period (\ref{firstp}). 
The wave function is of the form
\begin{eqnarray}
 \psi(T,t)= A(t)
\exp\left[-B(t)T^2\right] \ .
\label{evolve}
\end{eqnarray}
Solving the equation for $B(t)$ in $1/t$
expansion in the first period (\ref{firstp}), one obtains
\begin{eqnarray}
 B(t) = -\frac{v}{4\pi l_s^2 g_T}t + \frac{i}{4g_T} \frac{1}{t} 
+ \cdots \ .
\end{eqnarray}
Thus at $t=-t_0$, the initial condition for the second period is given
by
\begin{eqnarray}
 B(t=-t_0)  
= \frac{v}{4\pi l_s^2 g_T}t_0 - \frac{i}{4g_T} \frac{1}{t_0} \ . 
\label{ini}
\end{eqnarray}
In the second period, the evolution of the wave function
was determined in \cite{GP}. The solution is 
\begin{eqnarray}
 B(t^{\rm GP}) = \frac{1}{2a^2} \tan(\phi - i \omega t^{\rm GP})
\label{gps}
\end{eqnarray}
where $\phi$ is a parameter in the solution, and 
in correspondence to our language,
\begin{eqnarray}
 a^2 = \frac{g_T}{m_{\rm second}} \ , \quad 
\omega = m_{\rm second} \ .
\end{eqnarray}
$t^{\rm GP}$ is the time in the convention of \cite{GP} whose zero
$t_0^{\rm GP}$ is
different from that of our time as 
$t^{\rm GP}-t_0^{\rm GP}=t+t_0$.
We may derive this $t_0^{\rm GP}$ by 
comparing (\ref{ini}) and (\ref{gps}) at 
$t^{\rm GP}=t_0^{\rm GP}$. We obtain two equations
\begin{eqnarray}
 \frac{\sin 2\phi}{\cos 2\phi + \cosh 2\omega t_0^{\rm GP}}
= \sqrt{\frac{\theta}{s}} \ , \quad
\frac{\sin 2\phi}{\sinh 2\omega t_0^{\rm GP}} = \frac{2\theta}{v} \ .
\end{eqnarray}
For small $\theta$ and $v/\theta$, these can be solved to give
\begin{eqnarray}
 \sin 2\phi \simeq 2 \sqrt{\frac{\theta}{s}}, 
\quad \omega t_0^{\rm GP} \simeq \sqrt{\frac{v\theta}{s}} \ ,
\end{eqnarray}
which gives quite small $t_0^{\rm GP}$ ($\ll |t_0|$), thus 
we may use a relation 
\begin{eqnarray}
 t^{\rm GP} \simeq t+t_0 \ .
\end{eqnarray}

For large $\omega t$, a simple expression for the wave function was
given in \cite{GP}, 
\begin{eqnarray}
 |\psi|^2 = \sqrt{\frac{2\sin 2\phi}{a^2 \pi}} \exp
\left[-\omega t^{\rm GP}-2(\sin 2\phi) 
e^{-2\omega t^{\rm GP}} \frac{T^2}{a^2}\right] \ .
\end{eqnarray}
The wave function is still gaussian, due to which it is easy to finally 
evaluate the probability of having (\ref{taccondi}),
$T>\sqrt{\theta}/\sqrt{2\pi}l_s$ 
at $t=t_0$:
the reconnection probability is 
\begin{eqnarray}
 P = 2\int_{\sqrt{\theta}/\sqrt{2\pi}l_s}^\infty 
dT \; |\psi(T,t^{\rm GP}=2t_0)|^2 \ .
\end{eqnarray}
This is basically an error function whose asymptotic form is 
\begin{eqnarray}
\int_u^\infty \!\!ds\; e^{-s^2} = \frac{e^{-u^2}}{2u} + \cdots \ ,
\end{eqnarray}
thus we obtain
\begin{eqnarray}
 P \simeq \frac{\sqrt{g_s}}{2\pi^{3/4}\theta^{3/4}}
e^{2\sqrt{s}\theta/v}
\exp\left[-\frac{4\sqrt{\pi}\theta^{3/2}}{g_s}
e^{-4\sqrt{s}\theta/v}\right] \ .
\label{ourre}
\end{eqnarray}
This is the probability of the reconnection to occur for colliding
D-strings. 

Surprisingly, the result is very close to that of \cite{JJP}, a 
string worldsheet calculation,
\begin{eqnarray}
 P = \exp\left[
\left(
4-\frac{v}{2g_s}
\right)e^{-\pi\theta/v}
\right] \ .
\label{jjpre}
\end{eqnarray}
In fact, both have the ``nonperturbative'' factor in the exponent, 
$-e^{-\theta/v}/g_s$. This is quite interesting since the derivation of
these results are very different. The result (\ref{jjpre}) \cite{JJP}
has been derived by evaluating the probability of creation of pairs of
strings and anti-strings at the moment of the collision of the
D-strings, and looking at the force balance of the created string
junctions. Our result (\ref{ourre}) is based on the effective field
theory on the D-strings and the tachyon condensation. 

As noted in \cite{JJP}, when the string coupling $g_s$ becomes small,
the reconnection probability gets very small. Intuitively this is
obvious in our derivation, since small $g_{\rm s}$ corresponds to a
large $1/g$, that is, a particle with a large mass in the quantum
mechanics when $T$ is identified with the position of the
particle. Quantum evolution of a heavy particle is slow, and thus 
in our language correspondingly, the heaviness of the D-strings 
is reflected in the final form of the reconnection probability
(\ref{ourre}). 

A slight discrepancy between (\ref{ourre}) and (\ref{jjpre})
is found in the exponent, the coefficient of the nonperturbative factor
is $\theta^{3/2}$ in our case while $v$ in (\ref{jjpre}). This
difference might originate in the following point: in \cite{JJP} the
reconnection condition is derived by a requirement that kinks generated
by the creation of the F-strings between the D-strings cannot propagate
faster than the speed of light. In our case, the lower-bound for the
tachyon vev was given by the information of the geometrical shape of the
reconnected D-strings. These might be related to each other in a more
precise treatment of the quantum mechanics, since we have adopted a
crude approximation (\ref{firstp}), (\ref{secondp}) for the time
dependence of the potential which should have been dependent on the
velocity $v$. Furthermore, we have assumed that the tachyon vev is 
related directly only to the horizontal separation $\Delta$, which can
be improved. Another missing factor $4$ present in (\ref{jjpre}) might
appear in our calculation if we take into account fermions in this
system, since actually this $4$ came into the calculation in \cite{JJP}
from worldsheet fermions. 

\section{Summary and Discussions}
\label{sec5}

The intrinsic difference on collisions of the vortex strings in the
Abelian Higgs model and the D-strings in superstring theory is that the
reconnection occurs classically for the former while for the latter it
is merely a probabilistic event. This difference may be crucial for
recognizing if cosmic strings in future observations / early universe
scenarios are actually D-strings or not. In this paper we have clarified
theoretically the origin of this difference, by studying the
effective field theories on the vortex strings and on the D-strings. 
For the vortex strings, the theory is a sigma model whose target space
is a smeared cone geometry, and a Polyakov string moving smoothly in
this geometry describes the classical reconnection, showing its
inevitability. On the other hand, the D-string action allows a classical
solution of colliding D-strings passing without the reconnection, and
we performed a quantum time evolution of the tachyon wave function 
to give the reconnection probability. The reconnection is in fact 
the tachyon condensation. 

\begin{figure}[tp]
\begin{center}
\includegraphics[width=10cm]{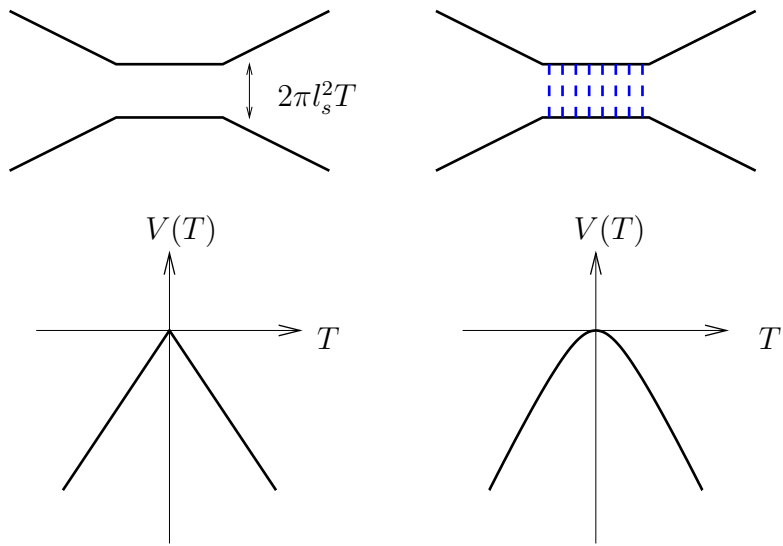}
\put(-186,166){ {$2\pi l_s^2 T$}}
\put(-171,74){ {$T$}}
\put(-236,116){ {$V(T)$}}
\put(0,74){{$T$}}
\put(-70,116){{$V(T)$}}
\end{center}
\begin{center}
\begin{minipage}{14cm}
\caption{A schematic picture of the bond connecting the reconnected
 D-strings \cite{TH2}. The upper-left figure shows strings
 reconnected hypothetically without the bond, while in the
 upper-right figure the bond is created in the actual
 reconnection of D-strings. If the bond was not created, the
 total energy of the D-strings would have  exhibited a non-zero
 one-point function at the zero separation (lower-left). This is not
 the case, because we have a tachyonic fluctuation which can be defined
 only if the top of the potential is a smooth maximum
 (lower-right). The difference in potential between the two graphs
 is supplemented by the existence of the bond energy.}
\label{bondfig}
\end{minipage}
\end{center}
\end{figure}

Then, why are the classical behaviors so different for these two
strings? Intuitively, it seems that the reconnection should always 
occur for collision of any slowly moving strings,  because the
reconnection decreases the total energy which is given by the tension
multiplied by the length of the strings. This should be a classical
understanding of the reconnection, and looks to be consistent with the
result for the vortex strings which always reconnect. Then why the
D-strings do not reconnect classically? A possible answer to this
question is found in \cite{TH2} 
where the energy of the created ``bond'' connecting the reconnected
D-strings is evaluated.\footnote{The identity of the bond might be a
bunch of fundamental strings, but classically the bond has no charge and 
thus is difficult to identify. But it should be something related to the
tachyon matter \cite{tacmat} because the reconnection is the tachyon
condensation \cite{TH2}.} 
This ``bond'' whose presence was suggested in \cite{sato, TH1}
should be there so that the classical reconnection is prevented.
The reconnection of D-strings is always accompanied by the creation of
the bonds, and resultantly, the total energy does not decrease for
infinitesimal reconnections \cite{TH2}. See Fig.~\ref{bondfig}.

Concerning the difference between the vortex strings and the 
D-strings, the following question\footnote{A related question was
given in \cite{blog}.} is worth answered: according to Sen's conjecture,
the D-strings are vortex-like topological defect in tachyon condensation
on a brane-antibrane system. The action of the system is given by the
tachyon field and gauge fields, and so it is almost the same as the
Abelian Higgs model. Thus any D-string can be regarded as a vortex
string in this sense.  
Then, why can those be different?\footnote{Concerning this question, it
would be interesting to pursue the connection proposed in \cite{Dvali}.}
Our answer to this question
is that actually the action of the brane-antibrane is {\it not} the
Abelian Higgs model, because there exists infinite number of massive
fields whose mass scale is the same as that of the tachyon field.
Only when one considers all the tower of the excitations simultaneously,
Sen's descent relation is properly realized: 
a D-string remains as a topological defect while the brane-antibrane
disappears. In fact, the disappearance of the brane-antibrane can be
shown only if one employs string field theories. Extracting just the
tachyon field and the gauge field does not provide the D-string
correctly. This is the very difference between ordinary field theories
such as the Abelian Higgs model and the string field theories. 
In ordinary field theories, any condensation of a field does not modify
the number of physical degrees of freedom, while in string field
theories this occurs, which is the significance of Sen's
conjectures.\footnote{By rotating the anti-brane against the brane, the
tachyon mass squared can reduce to nearly massless \cite{HN, TH1,ML},
but at the same time the massive fields comes down to massless, thus
treating solely the tachyon field is validated only for small
condensation of the tachyon.} 

In Section \ref{sec3}, we have parametrized the Moduli space 
(\ref{D-term}) by the coordinates (\ref{param}) in which $z$ specifies
the location of the vortex strings. As mentioned in the footnote there,
this $z$ is merely the approximate location of the strings because the
real part of the matrix $Z$ and the imaginary part are not
simultaneously diagonalizable for generic $z$. This is reflected in the
D-term condition (\ref{D-term}) as a non-commutativity: if we neglect 
the field $\psi$, the condition (\ref{D-term}) is just the Heisenberg
algebra and defines a non-commutative plane whose non-commutativity
parameter is the FI parameter $r$. Thus the location of the vortex
strings is ambiguous to the precision of ${\cal O}(r)$ in the $z$ space,
in terms of the vortex effective theory. It would be tempting to
calculate the actual energy distribution of the vortex strings in the 3
dimensional space, given the string configuration defined by $z$ in 
the vortex moduli space. This should be possible through the brane
configuration and string theory, because intrinsically this map is the
Seiberg-Witten map \cite{SW, OO}, or in other words, Matrix theory
multi-pole moments \cite{TV}, which describes a coupling of
non-commutative worldvolume configurations to bulk fields such as
gravity. In \cite{TH1, TH2, HO} explicit map has been evaluated for
various non-commutative configurations. To apply for our case of the
vortex strings, we  have to generalize the explicit Seiberg-Witten map
\cite{OO} to the situation where the D-branes end on some other D-branes
and hence additional fundamental fields are present.

\acknowledgments 

K.~H.~appreciated valuable discussions with E.~Copeland, T.~Kimura, 
M.~Nitta, H.~Ooguri, H.~Shimada and D.~Tong.
We would like to thank the Aspen Center for Physics for kind
hospitality and providing rich environment for stimulating discussions.
The research of A.H.~was supported in part by the CTP and LNS of
MIT and the U.S. Department of Energy under cooperative research 
agreement $\#$\ DE--FC02--94ER40818, and by BSF 
American--Israeli Bi--National Science Foundation. A.~H.~is also
indebted to a DOE OJI Award.
The work of K.~H.~was partly supported by the Grant-in-Aid for
Scientific Research (No.~12440060, 13135205, 15540256 and 15740143) from 
the Japan Ministry of Education, Science and Culture.


\newcommand{\J}[4]{{\sl #1} {\bf #2} (#3) #4}
\newcommand{\andJ}[3]{{\bf #1} (#2) #3}
\newcommand{\AP}{Ann.\ Phys.\ (N.Y.)}
\newcommand{\MPL}{Mod.\ Phys.\ Lett.}
\newcommand{\NP}{Nucl.\ Phys.}
\newcommand{\PL}{Phys.\ Lett.}
\newcommand{\PR}{ Phys.\ Rev.}
\newcommand{\PRL}{Phys.\ Rev.\ Lett.}
\newcommand{\PTP}{Prog.\ Theor.\ Phys.}
\newcommand{\hep}[1]{{\tt hep-th/{#1}}}

\end{document}